\documentclass[]{aa}
\usepackage{graphicx}
\usepackage{txfonts}
\begin{document}

\title{Two confirmed class I very low-mass objects in Taurus}

   \author{C. Dang-Duc
       \inst{1,3}
       \and
       N. Phan-Bao
       \inst{1,2}
       \and
       D.T. Dao-Van
       \inst{1}
       }

\offprints{N.~Phan-Bao}

\institute{Department of Physics, HCM International University-VNU, Block 6, 
           Linh Trung Ward, Thu Duc
           District, HCM city, Viet Nam. \\
           \email{pbngoc@hcmiu.edu.vn}
          \and
          Institute of Astronomy and Astrophysics, Academia Sinica. 
          PO Box 23-141, Taipei 106, Taiwan, ROC. 
          \and
      Faculty of Physics and Engineering Physics,
      HCM University of Science-VNU, 227 Nguyen Van Cu Street, 
      District 5, HCM city, Viet Nam.
          }

      \date{Received; accepted}

\abstract
{
[GKH94]~41 and IRAS~04191+1523B were previously identified to be proto-brown dwarf
candidates in Taurus. [GKH94]~41 was classified to be a class I object. The dereddened spectral energy distribution of the 
source was later found to be suggestive of a class II object.
IRAS 04191+1523B is a class I object that is the secondary
component of a binary.
}
{We determine the evolutionary stage of [GKH94]~41 and
estimate the final masses of the two proto-brown dwarf candidates.   
}
{We used archive millimeter observations to produce continuum maps and collected 
data from the literature to construct the
spectral energy distribution of the targets.
}
{Our continuum maps revealed that both [GKH94]~41 and IRAS~04191+1523B 
are surrounded by envelopes. This provides direct evidence that [GKH94]~41 is
a class I object, not class II, as previously classified.  
For IRAS~04191+1523B, our continuum map spatially resolved the binary. 
Our estimated final masses are below 49$^{+56}_{-27}$~$M_{\rm J}$ 
and 75$^{+40}_{-26}$~$M_{\rm J}$ 
for [GKH94]~41 and IRAS~04191+1523B, respectively. 
This indicates that both sources will likely become brown dwarfs or very low-mass stars.
Therefore, [GKH94]~41 and IRAS~04191+1523B are two new confirmed class I very low-mass objects. 
Their existence also supports the scenario that brown dwarfs have the same formation
stages as low-mass stars.
}
{}
 \keywords{brown dwarfs --- techniques: interferometric --- stars: formation --- stars: protostars}

\authorrunning{Dang-Duc et al.}
\titlerunning{Two class I VLM objects}

  \maketitle

\section{Introduction}
In 1995, the first brown dwarfs (BDs, 13$-$75~$M_{\rm J}$) were discovered 
(e.g., \citealt{rebolo}). Up to date, thousands of BDs have been
identified in star-forming regions and in the solar 
neighborhood. The coolest BDs (e.g., Y dwarfs) 
have temperatures as cool as the human body. 
BDs are thought to form by different mechanisms.
Two major models have been proposed for the BD formation.
In the starlike models, very low-mass (VLM) cores are produced by turbulent fragmentation \citep{padoan04} or gravitational fragmentation \citep{bonnell08} of 
molecular clouds.
In the ejection models, BDs are simply VLM embryos ejected from unstable multiple protostellar
systems by dynamical interaction (e.g., \citealt{bate02}). 

Observations of statistical properties of BDs such as 
initial mass function, velocity dispersion, multiplicity, 
accretion disks, and jets (see \citealt{luhman12} and references therein) 
have indicated that these statistical properties of BDs form a continuum with 
those of low-mass stars (see \citealt{whit} and references therein). 
This therefore strongly supports starlike models for BD formation. 
Recent observations of molecular outflows in 
class II BDs and VLM stars \citep{pb08, pb11, pb14} 
have shown that the molecular outflow process occurs 
in these VLM objects as a scaled-down version 
of that seen in low-mass stars. 
The authors also suggested that the associated accretion
process is possibly episodic and has a very low accretion rate. 
This may prevent
a VLM core that might be directly produced by a process of
turbulent fragmentation (\citealt{padoan04}) to accrete enough 
material to become a star. 
More observations of proto-BDs at earlier classes are clearly needed
to confirm these results. 

So far, a few candidates of class 0/I proto-BDs have been identified 
(e.g., \citealt{bourke06,lee13,palau14,morata}).
Of these candidates, only IC348-SMM2E and L328-IRS 
are class 0 objects with estimated final masses below 
the hydrogen-burning limit 
\citep{lee13,palau14}. 
\citet{andre12} identified a pre-BD with an estimated final mass
to be substellar. 
These sources are rare benchmarks of BDs in the protostellar
phase on which their origin can be studied. Their existence has also demonstrated that BDs 
and low-mass stars share the same manner of formation process.
Here, we report our study of two class I proto-BD candidates
in Taurus, [GKH94]~41 and IRAS~04191+1523B. 

\section{Targets}
We selected [GKH94]~41 (2MASS~J04194657+2712552) and IRAS~04191+1523B
(2MASS~J04220007+1530248) in the list of 352 members of 
Taurus from \citet{luhman10}. Within the class I objects, 
they are class I proto-BD candidates with spectral types later than M6. 

[GKH94]~41 was primarily identified as an embedded protostar in Taurus
based on its very red near-infrared colors and extended appearance \citep{gomez}. 
Its estimated spectral type of M7.5$\pm$1.5 implies that the central object
is substellar \citep{luhman09}. 
The flat spectral energy distribution (SED) of [GKH94]~41 in the infrared region
suggested the source to be in the class I or very young class II evolutionary stage
\citep{luhman09}.
Its dereddened SED was later found to be suggestive of 
a class II BD surrounded by a disk and not by an envelope \citep{furlan}.

IRAS~04191+1523B is the secondary component of a 6$''$.1 binary \citep{duchene}.
The spectral types that were estimated based on luminosity  
are K6$-$M3.5 and M6$-$M8 for 
IRAS~04191+1523A (2MASS~J04220043+1530212) and IRAS~04191+1523B, respectively \citep{luhman10}.
IRAS~04191+1523B is a class I object \citep{luhman10}.
Its spectral type in the range of M6$-$M8 indicates
that the central source is a substellar or VLM object.
In the following sections, we consider IRAS~04191+1523B to be an M7.0$\pm$1.0,
which is an average spectral type of M6 and M8.

\section{Millimeter continuum observations}
We searched the Combined Array for Research in 
Millimeter-wave Astronomy (CARMA) data archive for observations of
[GKH94]~41 and IRAS~04191+1523B. Both targets were observed 
on 2012 December 25 with CARMA (Project C1013). 
All six 10.4-meter, nine 6.1-meter antennas were 
operated in the C configuration 
at about 102 GHz (or $\sim$2.9 mm). Zenith opacities at 102 GHz were in the range of 
0.24$-$0.38. The CARMA correlator was set up in the continuum mode
of sixteen 500~MHz wide bands with 39 channels per band.
The quasars 3C~111 and 0510+180 were 
observed for gain, and 3C~84 for passband and flux calibration. 
The uncertainty in the absolute flux calibration is about 15\%.
We used the MIRIAD package adapted 
for the CARMA to reduce the continuum data. 
The synthesized beam sizes are about $2.19'' \times 1.53''$ and 
$1.90'' \times 1.59''$
(natural weighting) for [GKH94]~41 and IRAS~04191+1523B, respectively. 
The primary FWHM beam is about 81$''$ 
at the observed frequency. 
\section{Results and discussion}
\begin{figure}
   \centering
    \includegraphics[width=6cm,angle=-90]{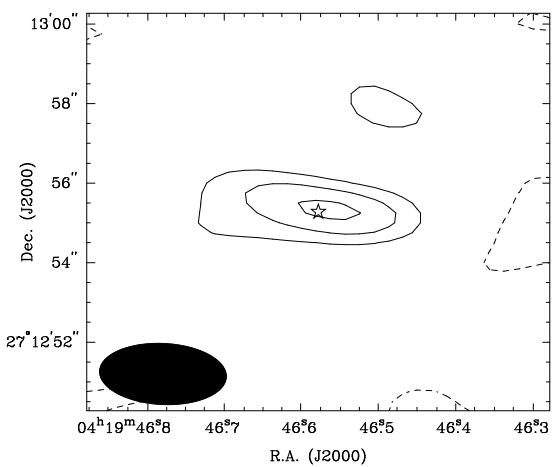}
    \caption{Continuum map at 2.9~mm of [GKH94]~41. 
The star symbol indicates the 2MASS near-infrared position of [GKH94]~41.
The contours are -3, 3, 5, and 7 times the rms of 
0.2~mJy~beam$^{-1}$. The synthesized beam is indicated in the bottom left corner.
}
\label{f1}
\end{figure}

The continuum emission from both targets is clearly detected. 
Figure~\ref{f1} shows the continuum map of [GKH94]~41. The position of 
the emission peak coincides with 
the 2MASS near-infrared position of the source.
This indicates that the continuum emission is associated with [GKH94]~41. 
The rms noise measured near the map center is about 0.2~mJy~beam$^{-1}$.
The integrated flux from [GKH94]~41 is
measured from a Gaussian fitting to be 2.5$\pm$0.2~mJy.
The deconvolved sizes of the emission that can be estimated from the
2D Gaussian fitting are listed in Table~\ref{info}. 
The deconvolved major axis of the emission region is about $2.9''$ or 
$\sim$400 AU in length at a distance of 140~pc (see Table~\ref{info}),
comparable to the typical size of disks 
associated with class I low-mass objects
(200$-$400~AU, e.g., \citealt{lee11}, \citealt{harsono}).
However, since we would expect that the disk size decreases with 
the mass of the central object and since the typical size of disks 
associated with class II BDs is significantly smaller 
(140$-$280 AU in diameter, e.g., \citealt{ricci14}), 
our large measured size suggests that 
the compact structure associated with [GKH94] 41 is dominated by an envelope and that this 
object is in the class I evolutionary stage.
The bolometric temperature estimated for [GKH94]~41 is $\sim$460~K, further
supporting that
the source is a late class I object (300$-$650~K, \citealt{enoch}).
We note that \citet{furlan} argued that the dereddened SED of
the source is similar to the SED of a typical T Tauri star with silicate emission
features at 10 and 20 $\mu$m and no ice absorption features.
The authors then concluded that [GKH94]~41 is a heavily extinguished 
class II object surrounded
by a disk and not by an envelope. 
A possible explanation for the \citet{furlan} result is that
the visual extinction value ($A_{\rm V}=27$) 
that was used in the dereddening of SED could
be overestimated (see Sect.~4.1).

\begin{figure}
   \centering
    \includegraphics[width=6cm,angle=-90]{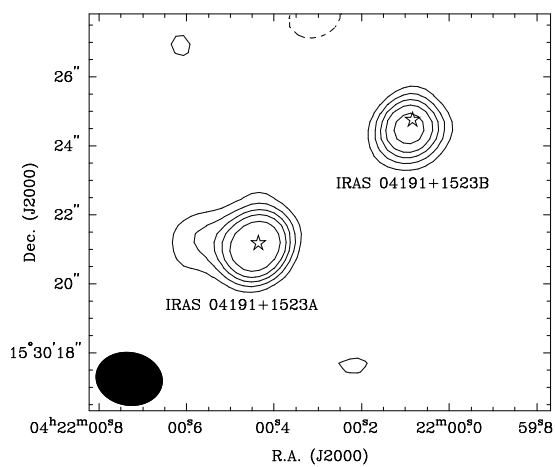}
    \caption{Continuum map at 2.9~mm of IRAS~04191+1523A
and IRAS~04191+1523B. Two components are spatially resolved with a separation of 6$''$.1.
The star symbols represent the 2MASS near-infrared positions of IRAS~04191+1523A and 
IRAS~04191+1523B.
The contours are -3, 3, 5, 7, 9, and 12 times the rms of 0.4~mJy~beam$^{-1}$. 
The synthesized beam is indicated in the bottom left corner.
}
\label{f2}
\end{figure}

For IRAS~04191+1523B (see Fig.~\ref{f2})
our map spatially resolves the binary into two components.
Both components are associated with envelopes.
The separation between the two peaks of emission from the envelopes is
about 6$''$.1, which is consistent with the previous measurement \citep{duchene}.
The rms noise measured near the map center is about 0.4~mJy~beam$^{-1}$.
The integrated fluxes are measured
from Gaussian fittings to be 9.5$\pm$0.4~mJy and 5.6$\pm$0.4~mJy
for IRAS~04191+1523A and IRAS~04191+1523B, respectively. 
The deconvolved sizes of the emission from IRAS~04191+1523A 
are listed in Table~\ref{info}. For IRAS~04191+1523B,
the Gaussian fitting is unable to deconvolve the emission, 
which appears to be a point source for 
the current synthesized beam size.

To determine whether [GKH94]~41 and IRAS~04191+1523B are class I proto-BDs, 
or in other words, whether these objects finally have substellar masses,
we estimate the upper limits to the final masses of the two objects
in the following sections.
\begin{table*}
  \caption{Estimated parameters of the 2.9 mm continuum sources associated with
[GKH94] 41, IRAS~04191+1523A, and IRAS~04191+1523B}
\label{info}
  $$
 \begin{tabular}{lcccccc}
   \hline 
   \hline
   \noalign{\smallskip}
Source           &  Deconvolved      & Deconvolved       &  P.A.           & Peak intensity       &  Flux       & $M_{\rm env}$$^{b}$  \\
                 & major axis ($''$) & minor axis ($''$) &  ($^{\circ}$)   & (mJy)                & (mJy)       & ($M_{\rm J}$)  \\
\hline
[GKH94]~41       & $2.9\pm0.1$      & $0.3\pm0.1$      & $83\pm7$        & $1.5\pm0.2$          & $2.5\pm0.2$ &     8          \\
IRAS~04191+1523A & $1.4\pm0.1$      & $0.3\pm0.1$      & $84\pm19$        & $7.0\pm1.0$         & $9.5\pm0.4$ &     31         \\
IRAS~04191+1523B & $1.8\pm0.1$$^{a}$& $1.6\pm0.1$$^{a}$& $-74\pm11$$^{a}$& $6.0\pm0.2$          & $5.6\pm0.4$ &     18         \\
   \hline
   \end{tabular}
   $$
\begin{list}{}{}
  \item[]
$^{a}$: Undeconvolved values; $^{b}$: Envelope masses (see Sect.~4 for further details).
\end{list}
\end{table*}
\subsection{[GKH94] 41: A class I proto-BD}
The mass that is added to the central object 
should be lower than the mass of material still available in the envelope
associated with the central object.
The upper limit to the final mass of a stellar object can therefore 
be estimated from the current mass of the central object 
and the mass of its envelope.

\begin{figure}
     \centering
    \includegraphics[width=6.0cm,angle=-90]{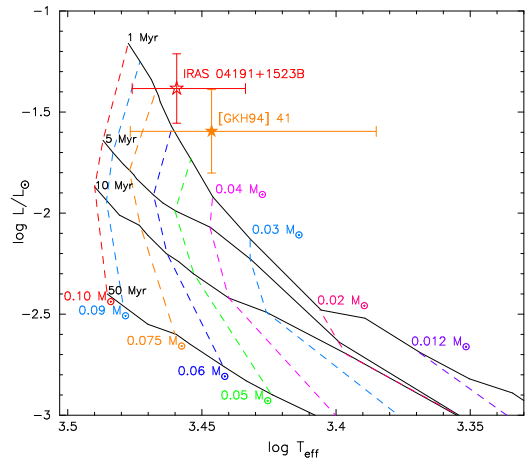}
    \caption{H-R diagram for [GKH94]~41 and IRAS~04191+1523B. Isochrones and mass tracks
from the theoretical evolutionary models of \citet{chabrier00} for $M\leq0.075$~$M_{\odot}$
and \citet{baraffe98} for $M>0.075$~$M_{\odot}$. 
The solid and open stars represent [GKH94]~41 (M7.5$\pm$1.5) 
and IRAS~04191+1523B (M7.0$\pm$1.0), respectively. 
The error bars reflect the uncertainties in the spectral type, 1.5 subclasses for 
[GKH94]~41 \citep{luhman09} and 1.0 subclass for IRAS~04191+1523B \citep{luhman10}.
}
\label{f3}
\end{figure}

First, to estimate the current mass of [GKH94]~41, 
we used the temperature versus mass relation 
as given in the DUSTY model \citep{chabrier00}. 
For a young object with spectral type M7.5, the temperature is 2795~K 
(see Table 8 in \citealt{luhman03}).
Because evolutionary models for VLM stars and BDs have high uncertainties at ages 
younger than 1~Myr \citep{baraffe02}, we  
assumed an age of 1~Myr for [GKH94]~41.
The temperature of 2795~K then corresponds to a mass of $\sim$41~$M_{\rm J}$. 
If the uncertainty of 1.5 subclasses in the spectral type is taken into account, 
a possible mass range for [GKH94]~41 is from 14 to 97~$M_{\rm J}$. 
We also calculated the luminosity of the source itself
(without any contribution from its surrounding envelope)
to constrain the mass range.
We first dereddened the 
2MASS $K$-band magnitude ($K_{\rm S} = 11.89$)
using the extinction law with $R_{\rm V}=3.1$ \citep{mathis} and
the visual extinction $A_{\rm V} = 4.0$ 
\citep{bulger14}. 
The $A_{\rm V}$ value was obtained from the best-fit 
atmospheric model (see \citealt{bulger14} and references therein)
over the 2MASS $JHK_{\rm S}$ bands. 
Using the bolometric correction
in $K$-band versus spectral type relation as given in \citet{filippazzo},
we derived a luminosity of 0.025~$L_{\odot}$.
Figure~\ref{f3} shows the luminosity versus temperature diagram for
[GKH94]~41 with the theoretical evolutionary models from \citet{chabrier00} and 
\citet{baraffe98}.
The location of [GKH94]~41 in the diagram  
indicates that the upper limit (including the uncertainties) to its current mass  
is just above the substellar
limit and below 0.09~$M_{\odot}$, 
which is consistent with 
the upper limit value of 97~$M_{\rm J}$  estimated above. 
We therefore conclude that the upper limit to the mass of [GKH94]~41 is 
in the range of 14$-$97~$M_{\rm J}$.

\begin{figure}
     \centering
    \includegraphics[width=6.0cm,angle=-90]{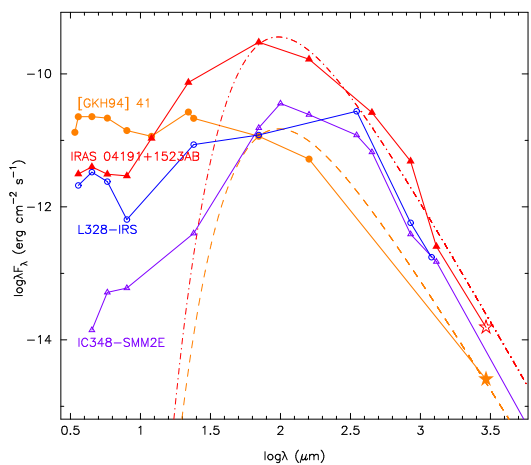}
    \caption{SEDs of the class I VLM objects ([GKH94]~41: brown line, 
IRAS~04191+1523AB: red line) 
and previously reported class 0 proto-BDs: 
L328-IRS \citep{lee13} (blue line) and 
IC348-SMM2E \citep{palau14} (violet line). 
For [GKH94]~41 and IRAS~04191+1523AB, infrared to submm data 
are taken from \citet{bulger14}
and references therein. The mm data measured in this paper are indicated by
the open and solid stars for IRAS~04191+1523AB and [GKH94]~41, respectively.
The brown dashed line 
shows the best fit for a modified blackbody of the dust envelope of
[GKH94]~41 and the red dash-dotted line for IRAS~04191+1523AB.
}
\label{f4}
\end{figure}

Second, we estimated the envelope mass of [GKH94]~41
based on the modeling of a modified blackbody 
with fluxes from 70~$\mu$m to mm wavelengths (Table~\ref{gkh}) as 
used to estimate the envelope mass of IC348-SMM2E (see \citealt{palau14}).
We obtain the dust temperature $T_{\rm d}$, dust emissivity $\beta,$
and envelope mass $M_{\rm env}$ by searching these three parameters
minimizing the $\chi^2$ in the following ranges:  
$T_{\rm d}$ from 10~K to 40~K, $\beta$ from 0.1 to 1.8 and 
$M_{\rm env}$ from 1~$M_{\rm J}$ to 60~$M_{\rm J}$.
We assumed a dust opacity coefficient 
dependence on wavelength $\lambda$ with 
$\kappa_{\rm \lambda} \propto \lambda^{-\beta}$, 
$\kappa_{\rm 870\mu m} = 0.0175$~cm$^{2}$~g$^{-1}$
(see \citealt{palau14} and references therein) 
and a gas-to-dust ratio of 100.
The best fit (Fig.~\ref{f4}) then gives  
$T_{\rm d}$ = 34~K, $M_{\rm env}$ = 2~$M_{\rm J}$, and $\beta$ = 0.4.
The obtained dust emissivity index value is significantly lower than 
the typical value of about 1.4
for class I objects in Taurus \citep{chandler}. This is probably because
the best fit is obtained from only three available data points.
If we take $\beta$ = 1.4 and $T_{\rm d}$ = 34~K, 
we obtain an envelope mass of 8~$M_{\rm J}$, 
a factor of 4 higher than the value estimated from the fitting. 

We therefore take the final mass of [GKH94]~41 to be
49~$M_{\rm J}$, which the sum of the current mass (41~$M_{\rm J}$)
and the envelope mass (8~$M_{\rm J}$). 
If the uncertainty in the spectral type 
is taken into account, this upper limit will be 49$^{+56}_{-27}$~$M_{\rm J}$. 
As the outflow process also ejects accreting material and dissipates material in the envelope, 
the final mass of [GKH94]~41 thus will very likely be below the substellar limit.

\begin{table}
  \caption{Photometry for [GKH94]~41}
\label{gkh}
  $$
 \begin{tabular}{ccccc}
   \hline 
   \hline
   \noalign{\smallskip}
Wavelength & Flux  & Error & References \\
($\mu$m)   & (mJy) & (mJy) &            \\
\hline
70         & 269   & 5     & 1          \\
160        & 279   & 66    & 1          \\
2940       & 2.5   & 0.2   & this paper \\
   \hline
   \end{tabular}
   $$
\begin{list}{}{}
  \item[]
References: (1) \citet{bulger14}.
\end{list}
\end{table}
\begin{table}
  \caption{Photometry for IRAS~04191+1523AB}
\label{iras}
  $$
 \begin{tabular}{ccccc}
   \hline 
   \hline
   \noalign{\smallskip}
Wavelength & Flux  & Error & References \\
($\mu$m)   & (mJy) & (mJy) &            \\
\hline
70         & 7002  & 13    &  1          \\
160        & 8884  & 232   &  1          \\
450        & 3940  & 220   &  2          \\
850        & 1380  & 20    &  2          \\
1300       & 110   & 7     &  3          \\
2940       & 9.5$^{a}$ & 0.4$^{a}$ & this paper \\
           & 5.6$^{b}$ & 0.4$^{b}$ & this paper \\
   \hline
   \end{tabular}
   $$
\begin{list}{}{}
  \item[]
References: (1) \citet{bulger14}; (2) \citet{francesco08}; (3) \citet{motte01}.
$^{a}$: the flux and its error for component IRAS~04191+1523A.
$^{b}$: the flux and its error for component IRAS~04191+1523B.
\end{list}
\end{table}
\subsection{IRAS~04191+1523B: A class I proto-BD or VLM star}
To estimate the upper limit to the final mass of IRAS~04191+1523B, we
followed the same steps as for [GKH94]~41.
A spectral type of M7.0 corresponds to
a temperature of 2880~K \citep{luhman03}.
Using the DUSTY model for an age of 1~Myr, 
we derive a mass of $\sim$57~$M_{\rm J}$ for the source. 
If we take the uncertainty of 1.0 subclass in the spectral type into
account, a possible mass range is
31$-$97~$M_{\rm J}$ for the object.
To calculate the luminosity of the source,
we used $A_{\rm V} = 34.4$ 
\citep{bulger14} and then derived a luminosity 
of 0.041~$L_{\odot}$.
The location of IRAS~04191+1523B in the luminosity
versus temperature diagram 
(Fig.~\ref{f3})
also indicates that the mass (including the uncertainties) of
the central source is below $\sim$0.09~$M_{\odot}$, which 
is consistent with the mass range derived from its spectral type.
The upper limit to the current mass of 
IRAS~04191+1523B is therefore in
the range of 31$-$97~$M_{\rm J}$. 
 
The fluxes at infrared and submm wavelengths are unresolved for 
components A and B. 
We thus used fluxes from 70~$\mu$m to mm wavelengths (Table~\ref{iras}) 
from the two components to estimate the total mass of the envelopes.
We applied the same method as for [GKH94]~41. 
The best fit (Fig.~\ref{f4}) gives 
$T_{\rm d}$ = 32~K, $M_{\rm env}$ = 33~$M_{\rm J}$, and $\beta$ = 0.7 for 
IRAS~04191+1523AB.
The envelope mass of each component could also be estimated directly from fluxes 
at mm wavelengths. 
If we take $\beta$ = 1.4 \citep{chandler} and $T_{\rm d}$ = 32~K, 
we obtain the envelope masses of about 31~$M_{\rm J}$ and 18~$M_{\rm J}$ for
IRAS~04191+1523A and IRAS~04191+1523B, respectively. The total mass
is now higher by a
factor of $\sim$1.5  than the value estimated from the fitting.   

With the current mass of 57~$M_{\rm J}$ and the envelope mass of 18~$M_{\rm J}$, 
the upper limit to the final mass of IRAS~04191+1523B is thus 75~$M_{\rm J}$. 
If we include the uncertainty in its spectral type, the upper limit
is 75$^{+40}_{-26}$~$M_{\rm J}$. This upper limit also indicates that
IRAS~04191+1523B will likely become a BD
or a VLM star with a mass just above the substellar limit.
\section{Conclusion}

We presented the detection of envelopes associated with the
two class I VLM objects [GKH94]~41 and IRAS~04191+1523B.
Our estimated final masses demonstrate that [GKH94]~41 and IRAS~04191+1523B
will likely become BDs or VLM stars. 
The existence of these class I VLM objects strongly supports
the scenario that BDs and VLM stars have the same formation processes as low-mass stars.

\begin{acknowledgements}
This research is funded by Vietnam National Foundation for Science
and Technology Development (NAFOSTED) under grant number 103.08-2013.21. 
We would like to thank the referee for useful comments that significantly improved the paper. 
Support for CARMA construction was derived from the Gordon and 
Betty Moore Foundation, the Kenneth T. and Eileen L. Norris Foundation, 
the James S. McDonnell Foundation, the Associates of the 
California Institute of Technology, the University of Chicago, 
the states of California, Illinois, and Maryland, and 
the National Science Foundation. Ongoing CARMA development and 
operations are supported by the National Science Foundation 
under a cooperative agreement, and by the CARMA partner universities.
\end{acknowledgements}


\begin{thebibliography}{}

\bibitem[Andr\'e et al.(2012)]{andre12}    
  Andr\'e, P., Ward-Thompson, D., \& Greaves, J. 2012, Science, 337, 69

\bibitem[Baraffe et al.(1998)]{baraffe98}
  Baraffe, I., Chabrier, G.,  Allard, F., \& Hauschildt, P. H. 1998, \aap, 337, 403

\bibitem[Baraffe et al.(2002)]{baraffe02}
  Baraffe, I., Chabrier, G.,  Allard, F., \& Hauschildt, P. H. 2002, \aap, 382, 563

\bibitem[Bate et al.(2002)]{bate02}
  Bate, M. R., Bonnell, I. A., \& Bromm, V. 2002, MNRAS, 332, L65

\bibitem[Bonnell et al.(2008)]{bonnell08} 
  Bonnell, I. A., Clark, P., \& Bate, M.R. 2008, MNRAS, 389, 1556

\bibitem[Bourke et al.(2006)]{bourke06}
  Bourke, T. L., Myers, P. C., Evans II, N. J., et al. 2006, \apj, 649, L37

\bibitem[Bulger et al.(2014)]{bulger14} 
  Bulger, J., Patience, J., Ward-Duong, K., Pinte, C., Bouy, H., M\'enard, F., \& Monin, J.-L. 2014, A\&A, 570, A29

\bibitem[Chabrier et al.(2000)]{chabrier00}
  Chabrier, G., Baraffe, I., Allard, F., \& Hauschildt, P. H. 2000, \apj, 542, 464

\bibitem[Chandler et al.(1998)]{chandler}
  Chandler, C. J., Barsony, M., \& Moore, T. J. T. 1998, MNRAS, 299, 789

\bibitem[Duch\^ene et al.(2004)]{duchene}
  Duch\^ene, G., Bouvier, J., Bontemps, S., Andr\'e, P., \& Motte, F. 2004, \aap, 427, 651

\bibitem[Filippazzo et al.(2015)]{filippazzo}
  Filippazzo, J. C., Rice, E. L., Faherty, J., Cruz, K. L., Van Gordon, M. M., \& Looper, D. L.
  2015, \apj, 810, 158

\bibitem[Enoch et al.(2009)]{enoch} 
  Enoch, M. L., Evans, N. J. II, Sargent, A. I., \& Glenn, J. 2009, \apj, 692, 973

\bibitem[Francesco et al.(2008)]{francesco08}
  Francesco, J. D., Johnstone, D., Kirk, H., MacKenzie, T., \& Ledwosinska, E.
  2008, \apjs, 175, 277

\bibitem[Furlan et al.(2011)]{furlan} 
  Furlan, E., Luhman, K. L., Espaillat, C., et al. 2011, \apjs, 195, 3

\bibitem[Gomez et al.(1994)]{gomez} 
  Gomez, M., Kenyon, S. J., \& Hartmann, L. 1994, \aj, 107, 1850

\bibitem[Harsono et al.(2014)]{harsono} 
  Harsono, D., J{\o}rgensen, J. K., van Dishoeck, E. F., Hogerheijde, M. R., Bruderer, S., 
  Persson, M. V., \& Mottram, J. C. 2014, \aap, 562, A77

\bibitem[Lee(2011)]{lee11} 
  Lee, C.-F. 2011, ApJ, 741, 62

\bibitem[Lee et al.(2013)]{lee13} 
  Lee, C. W., Kim, M.-R., Kim, G., Saito, M., Myers, P. C., \& Kurono, Y. 2013, ApJ, 777, 50

\bibitem[Luhman et al.(2003)]{luhman03} 
  Luhman, K. L., Stauffer, J. R., Muench, A. A., Rieke, G. H., Lada, E. A., Bouvier, J., 
  \& Lada, C. J. 2003, \apj, 593, 1093

\bibitem[Luhman et al.(2009)]{luhman09} 
  Luhman, K. L., Mamajek, E. E., Allen, P. R., \& Cruz, K. L. 2009, ApJ, 703, 399

\bibitem[Luhman et al.(2010)]{luhman10} 
  Luhman, K. L., Allen, P. R., Espaillat, C., Hartmann, L., \& Calvet, N. 2010, \apjs, 186, 111

\bibitem[Luhman(2012)]{luhman12} 
  Luhman, K. L. 2012, ARA\&A, 50, 65

\bibitem[Mathis(1990)]{mathis} 
  Mathis, J. S. 1990, ARA\&A, 28, 37

\bibitem[Morata et al.(2015)]{morata} 
  Morata, O., Palau, A., Gonz\'alez, R. F., et al. 2015, ApJ, 807, 55

\bibitem[Motte \& Andr\'e(2001)]{motte01}
  Motte, F., \& Andr\'e, P. 2001, \aap, 365, 440

\bibitem[Padoan \& Nordlund(2004)]{padoan04} 
   Padoan, P., \& Nordlund, \AA. 2004, ApJ, 617, 559

\bibitem[Palau et al.(2014)]{palau14} 
   Palau, A., Zapata, L. A., Rodr\'{\i}guez, L. F., et al. 2014, MNRAS, 444, 833

\bibitem[Phan-Bao et al.(2008)]{pb08} 
   Phan-Bao, N., Riaz, B., Lee, C.-F., et al. 2008, \apjl, 689, L141

\bibitem[Phan-Bao et al.(2011)]{pb11} 
   Phan-Bao, N., Lee, C.-F., Ho, P. T. P., \& Tang, Y.-W. 2011, ApJ, 735, 14

\bibitem[Phan-Bao et al.(2014)]{pb14} 
  Phan-Bao, N., Lee, C.-F., Ho, P. T. P., Dang-Duc, C., \& Li, D. 2014, ApJ, 795, 70

\bibitem[Rebolo et al.(1995)]{rebolo}
  Rebolo, R., Zapatero Osorio, M. R., \& Mart{\'{i}}n, E. L. 1995, Nature, 377, 129

\bibitem[Ricci et al.(2014)]{ricci14} 
  Ricci, L., Testi, L., Natta, A., Scholz, A., de Gregorio-Monsalvo, I., \& Isella, A. 
  2014, \apj, 791, 20

\bibitem[Whitworth et al.(2007)]{whit} 
  Whitworth, A., Bate, M.~R., Nordlund, \AA.,  2007, in Protostars and Planets V, 
ed. B. Reipurth, D. Jewitt, \& K. Keil (Tucson, AZ: Univ. Arizona Press), 459

\end{thebibliography}
\end{document}